\documentclass[letterpaper, 10 pt, conference]{ieeeconf}  

\IEEEoverridecommandlockouts                              

\usepackage{cite}
\usepackage{amsmath,amssymb,amsfonts}
\usepackage{algorithmic}
\usepackage{graphicx}
\usepackage{textcomp}
\usepackage{xcolor}
\usepackage{url}
\usepackage{balance}

\usepackage[acronym]{glossaries}
\usepackage{glossaries-prefix}
\usepackage{url}
\usepackage{tabularx}
\usepackage{multirow}
\usepackage{microtype}
\usepackage{subfig}
\usepackage{booktabs}

\let\labelindent\relax
\usepackage[shortlabels]{enumitem}

\usepackage{tikz}
\usepackage{textcomp}
\usepackage{hyperref}
\usepackage{lipsum}

\newcommand{\cready}[1]{#1}

\definecolor{red3}{RGB}{136,0,3}
\definecolor{comment}{named}{red3}

\newboolean{showcomments}
\setboolean{showcomments}{true} 
\ifthenelse{\boolean{showcomments}}
  {
		\newcommand{\nbb}[2]{
		\fcolorbox{black}{yellow}{\bfseries\sffamily\scriptsize#1}
		{\sf$\blacktriangleright$\textcolor{blue}{\textit{#2}}$\blacktriangleleft$}
		}
		
		\newcommand{\remarks}[1]{\color{red}[#1]\color{black}}

		\newcommand{\del}[1]{\textcolor{red}{\sout{#1}}} 
  }
  {
		\newcommand{\nbb}[2]{}
		\newcommand{\remarks}[1]{}

		\newcommand{\del}[1]{} 
  }

\newcommand{\subrange}{sub-parameter range~}
\newcommand{\subrangepl}{sub-parameter ranges~}
\newcommand{\Subrange}{Sub-Parameter Range~}

\newacronym[
    prefixfirst={an\ },
    prefix={an\ }
    ]{ACC}{ACC}{Adaptive Cruise Control}
\newacronym[
    prefixfirst={an\ },
    prefix={an\ }
    ]{ADAS}{ADAS}{Advanced Driver Assistance System}
\newacronym[
    prefixfirst={an\ },
    prefix={an\ }
    ]{ADS}{ADS}{Automated Driving System}
\newacronym{CAD}{CAD}{Computer-Aided Design}
\newacronym[
    prefixfirst={an\ },
    prefix={an\ }
    ]{FM}{FM}{Feature Model}
\newacronym[
    prefixfirst={an\ },
    prefix={an\ }
    ]{ODD}{ODD}{Operational Design Domain}
\newacronym{SOTIF}{SOTIF}{Safety of the Intended Functionality}
\newacronym[
    prefixfirst={an\ },
    prefix={an\ }
    ]{SUT}{SUT}{System Under Test}
\newacronym[
    prefixfirst={an\ },
    prefix={an\ }
    ]{SRC}{SRC}{Safety Relevance Controller}

\newcommand\copyrighttext{%
      \footnotesize \textcopyright 2023 IEEE. Personal use of this material is permitted.
      Permission from IEEE must be obtained for all other uses, in any current or future
      media, including reprinting/republishing this material for advertising or promotional
      purposes, creating new collective works, for resale or redistribution to servers or
      lists, or reuse of any copyrighted component of this work in other works.
      }
      
\newcommand{\copyrightnotice}{%
    \begin{tikzpicture}[remember picture, overlay]
        \node[anchor=south,yshift=10pt] at (current page.south) {\fbox{\parbox{\dimexpr\textwidth-\fboxsep-\fboxrule\relax}{\copyrighttext}}};
    \end{tikzpicture}%
}

\title{\LARGE \bf
SOTIF-Compliant Scenario Generation Using \\ Semi-Concrete Scenarios and Parameter Sampling
}

\author{Lukas Birkemeyer$^{1}$, Julian Fuchs$^{2}$, Alessio Gambi$^{3}$ and Ina Schaefer$^{4}$
\thanks{$^{1}$Lukas Birkemeyer is with Technical University Braunschweig, Braunschweig, Germany
        {\tt\small l.birkemeyer@tu-braunschweig.de}}%
\thanks{$^{2}$Julian Fuchs is with FZI Forschungszentrum Informatik, Karlsruhe, Germany
        {\tt\small fuchs@fzi.de}}%
\thanks{$^{3}$Alessio Gambi is with IMC University of Applied Sciences, Krems, Austria
        {\tt\small alessio.gambi@fh-krems.ac.at}}%
\thanks{$^{4}$Ina Schaefer is with Karlsruhe Institute of Technology, Karlsruhe, Germany
        {\tt\small ina.schaefer@kit.edu}}%
}

\def\BibTeX{{\rm B\kern-.05em{\sc i\kern-.025em b}\kern-.08em
    T\kern-.1667em\lower.7ex\hbox{E}\kern-.125emX}}

\begin{document}

\maketitle
\thispagestyle{empty}
\pagestyle{empty}

\begin{abstract}

    The SOTIF standard (ISO 21448) requires scenario-based testing to verify and validate Advanced Driver Assistance Systems and Automated Driving Systems but does not suggest any practical way to do so effectively and efficiently. 
    Existing scenario generation approaches either focus on exploring or exploiting the scenario space. This generally leads to test suites that cover many known cases but potentially miss edge cases or focused test suites that are effective but also contain less diverse scenarios.
    To generate SOTIF-compliant test 
    suites that achieve higher coverage and find more faults, this paper proposes semi-concrete scenarios and combines them with parameter sampling to adequately balance scenario space exploration and exploitation. Semi-concrete scenarios enable combinatorial scenario generation techniques that systematically explore the scenario space, while parameter sampling allows for the exploitation of continuous parameters.
    Our experimental results show that the proposed concept can generate more effective test suites than state-of-the-art coverage-based sampling.
    Moreover, our results show that including a feedback mechanism to drive parameter sampling further increases test suites' effectiveness.
    
\end{abstract}

\section{Introduction}

    \copyrightnotice
    The \gls{SOTIF}-standard (ISO~21448) \cite{iso21448}
    requires scenario-based testing to validate \glspl{ADAS} and \glspl{ADS}.
    In scenario-based testing, scenarios precisely describe relevant environmental elements to include in the testing process and the \gls{SUT}'s initial state.
    \textit{Logical scenarios} describe the main semantics of the scenarios and include parameters that influence their instantiation~\cite{steimle2021toward}. For example, a hypothetical logical scenario for testing an \gls{ACC} might occur on a highway; the ego vehicle travels at a given \emph{speed} ($v_{ego}$) and approaches \emph{another vehicle}.
    Watanabe~\cite{watanabe2022methodik} distinguishes between \textit{continuous} parameters (e.g., vehicle speed) and \textit{discrete} parameters (e.g., vehicle model). Continuous parameters are usually defined in terms of ranges (e.g., $v_{ego}=[120,150]~km/h$).
    Scenario-based testing requires instantiating logical scenarios in \textit{concrete scenarios}, i.e., test cases, by assigning specific values to continuous and discrete parameters~\cite{bagschik2018wissensbasierte}. For instance, in a concrete scenario, the ego-vehicle travels at $123~km/h$ on a highway and approaches a \textit{VW Beetle}.
    The set of all the generated concrete scenarios forms a test suite (i.e., scenario suite).
    
    The \gls{SOTIF} standard requires scenario-based testing to ensure that \pgls{ADAS}/\gls{ADS} operates as intended within a specified \gls{ODD}~\cite{iso21448}, i.e., a set of well-defined execution conditions and their possible interplay. However, \gls{SOTIF} does not suggest any concrete strategy to select discrete parameters and sample continuous ones; thus, the challenge of instantiating logical scenarios into concrete ones remains open.
    \cready{According to Birkemeyer et al.~\cite{birkemeyer2023literature}, e}xisting research on parameter selection and scenario generation includes:
    \emph{data-driven} techniques that select parameters based on real-world data instances~\cite{gambi2019generating, montanari2021maneuver} or distributions~\cite{de2017assessment, krajewski2019beziervae};   
    \textit{optimization} techniques that exploit parameter spaces using search-based methods~\cite{gambi2019automatically, gambi2019asfault, abdessalem2018testing, althoff2018automatic, ghodsi2021generating, tuncali2019rapidly, ding2021multimodal} or machine learning~\cite{baumann2021automatic}; and, \textit{combinatorial} techniques that explore scenario spaces by systematically combining atomic scenario elements~\cite{birkemeyer2022feature, kluck2023empirical}.
    %
    \gls{SOTIF} requires that scenario suites represent the \gls{ODD}~\cite{iso21448}. Thus, we argue that although data-driven scenario generation can generate realistic scenarios and optimization-based scenario generation can generate critical scenarios, only combinatorial scenario generation, which systematically covers all interactions of $t$ atomic scenario elements, has the potential to generate SOTIF-compliant test suites.
    However, \gls{SOTIF} also requires that the set of scenarios that lead to unsafe behavior of the \gls{SUT} (i.e., critical scenarios) becomes minimal~\cite{iso21448}. Since existing combinatorial scenario generation techniques cannot provide this property~\cite{kluck2023empirical}, they must be extended to balance scenario space exploration vs. exploitation.

    To solve this issue, in this paper, we propose to generate effective, \gls{SOTIF}-compliant test suites by combining combinatorial scenario generation techniques and parameter sampling.
    The key enabler of the proposed approach is semi-concrete scenarios that conceptually sit between logical and concrete scenarios. Semi-concrete scenarios enable covering discrete parameters (exploration) while optimizing continuous parameters (exploitation).
    We empirically evaluate the proposed approach by assessing the effectiveness of the test suites it generates, and we observed these test suites find more faults in the \gls{SUT} than those generated by existing combinatorial scenario generation techniques. 
        
    This paper makes 
    the following contributions:
    \begin{enumerate}
        \item A scenario generation technique based on the novel concept of semi-concrete scenarios that combines combinatorial testing and parameter sampling to generate effective test suites.
        \item An empirical evaluation of the effectiveness of test suites generated from semi-concrete scenarios using combinatorial generation and parameter sampling.
    \end{enumerate}
        
\section{Variability Modeling Techniques}

    This section provides fundamental knowledge regarding feature modeling and sampling strategies to make the paper self-contained.
        
\begin{figure}
                \centering
                \includegraphics[width=\linewidth]{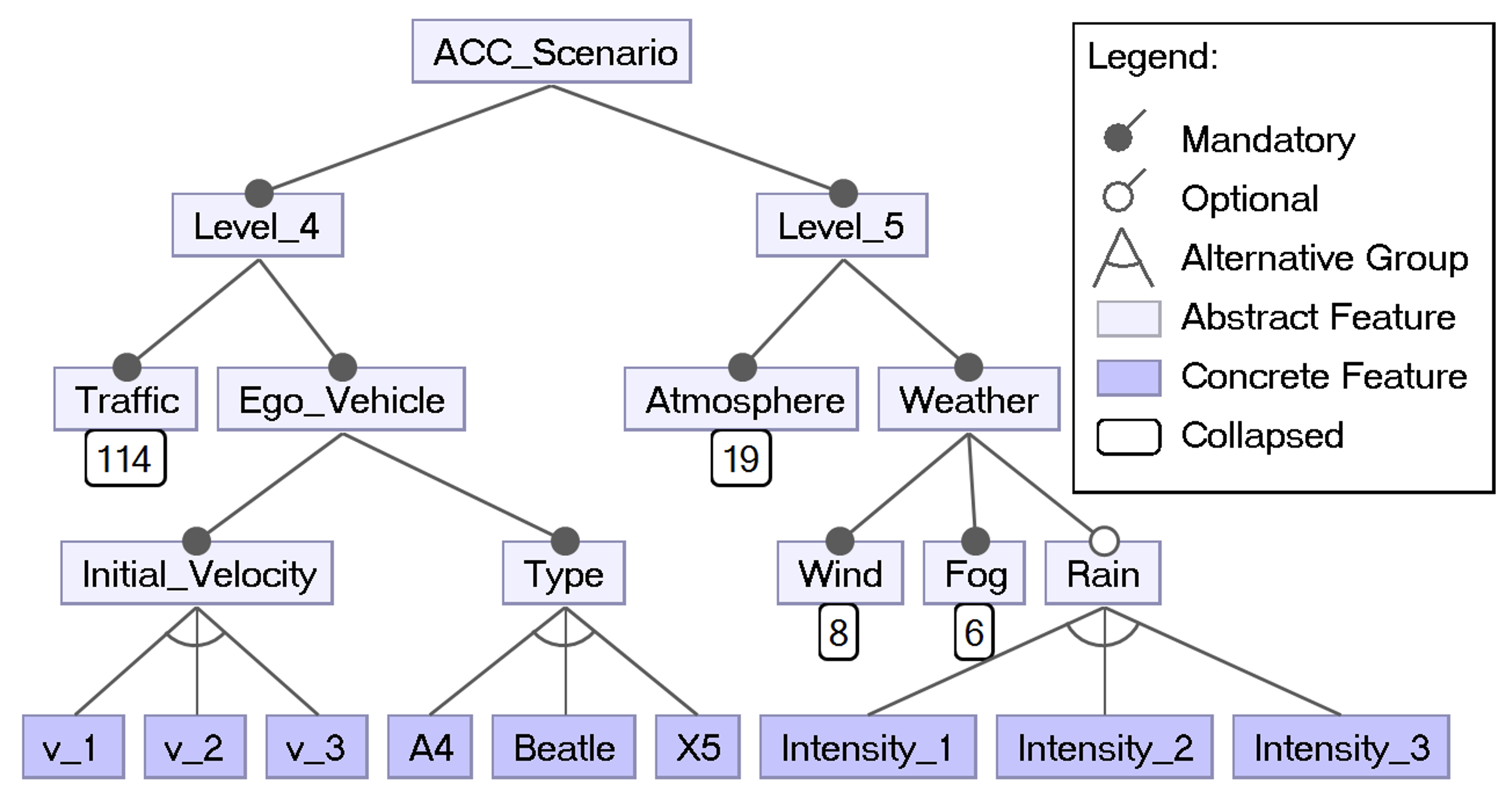}
                \caption{
                    Excerpt of the feature model 
                    proposed in~\cite{birkemeyer2022feature}. The \textit{initial velocity} and \textit{type} of the ego-vehicle are mandatory features, while the feature \textit{rain} is optional.
                }
                \label{fig:FM_excerpt}
            \end{figure}

        \paragraph{Feature Modeling}
            Variability modeling techniques are common methods to model highly configurable systems in software engineering \cite{apel2016feature, kastner2009featureide}. Configuration options are modeled as \textit{features} in a \textit{feature model}. A feature is a binary, user-visible system configuration option (e.g., the color or type of a vehicle). A feature model has a tree structure and describes characteristics of features (optional/mandatory) and dependencies between features (alternative/AND/OR as well as parent/child - relation). In Figure~\ref{fig:FM_excerpt}, we present an excerpt of a feature model that represents scenarios for scenario-based testing inspired by Birkemeyer et al.~\cite{birkemeyer2022feature}. The feature model is structured according to the six-scenario levels proposed by Scholtes et al.~\cite{scholtes20216} and covers atomic scenario elements. An atomic scenario element is a concrete entity of a scenario such as rain with the intensity of $1 mm/h$ or the initial velocity of a vehicle $v_{ego}=25 km/h$. Thus, the scenario feature model introduced by Birkemeyer et al.~\cite{birkemeyer2022feature} covers the space of possible scenarios.

        \paragraph{Sampling Strategies}

            Due to the combinatorial explosion, the number of valid configurations represented by a feature model becomes extremely large \cite{pett2019product, varshosaz2018classification}; hence, testing all scenarios is practically infeasible and sampling strategies to select a subset of all valid configurations~\cite{johansen2011properties, johansen2012algorithm, krieter2020yasa} must be used to generate concrete scenarios.
            Selecting a set of configurations that is representative of the overall configuration space makes it possible to derive sound assumptions about the adequacy of the generated test suites.
            Coverage-based sampling strategies select configurations so that each feature or interaction of $t$ features is covered at least once in the representative subset. Considering the scenario feature model, coverage-based sampling explores the scenario space by ensuring that the resulting scenario suite contains each combination of $t$ atomic scenario elements. State-of-the-art algorithms for coverage-based sampling are Chvatal~\cite{johansen2011properties}, ICPL~\cite{johansen2012algorithm}, and YASA~\cite{krieter2020yasa}.

\section{Balancing Exploration and Exploitation}

    \begin{figure}
        \centering
        \includegraphics[width=\linewidth]{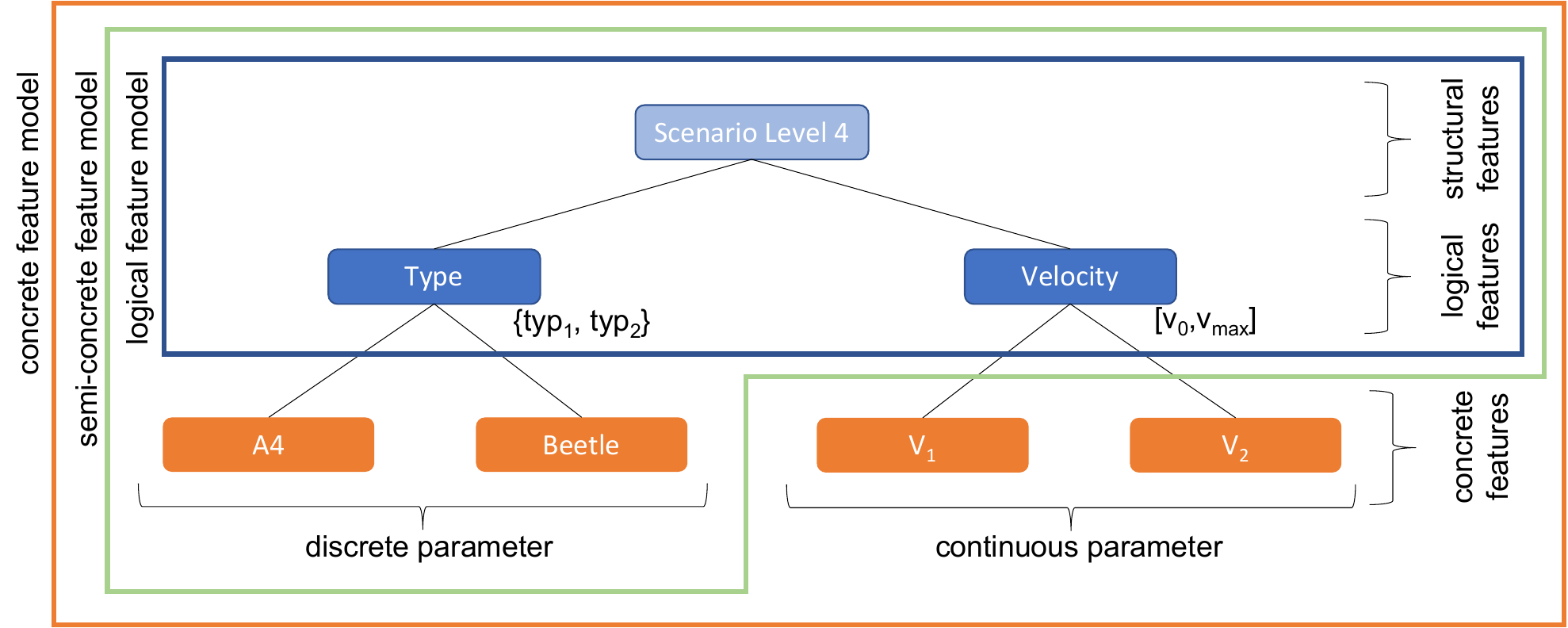}
        \caption{Hybrid scenario feature model 
        to represent logical (blue), semi-concrete (green), and concrete (orange) scenarios. The logical scenario feature model is part of the semi-concrete and part of the concrete scenario feature model.}
        \label{fig:logical_concrete_FM}
    \end{figure}

    This paper aims to establish a novel concept for scenario generation that allows balancing exploration and exploitation of scenario spaces to improve the test effectiveness of \gls{SOTIF}-compliant scenario-based testing. Klück et al.~\cite{kluck2023empirical} state that combinatorial scenario generation (exploration) is more effective in detecting fault than optimization-based scenario generation (exploitation).
    Thus, we use combinatorial scenario generation as a starting point.
    However, we observed that standard combinatorial scenario generation, which uses equivalence classes, i.e., predefined value assignments, to discretize continuous parameters, is sub-optimal. 
    The following example demonstrates that equivalence classes do not properly represent a continuous parameter space because one parameter might occur in different scenarios, possibly requiring it to take different values. 
    In this example, we consider only the initial velocity of the ego vehicle $v$. 
    On a straight road (\textit{scenario-1}), $v_1$ represents an equivalence class. If we, however, replace the straight road with a sharp turn (\textit{scenario-2}), $v_2$ represents an equivalence class that is different from the equivalence class defined for \textit{scenario-1} ($v_1 \neq v_2$).
    To sum up, existing combinatorial scenario generation leads to effective scenario suites but has issues in discretizing continuous parameters.
    Optimization techniques, in contrast, are explicitly designed to select concrete values from continuous parameters that optimize a fitness function \cite{papageorgiou2015optimierung}. Alternative parameter sampling strategies select parameters randomly~\cite{kluck2023empirical}, equidistantly~\cite{skruch2021approach}, or based on probability density functions \cite{nakamura2022defining}.
    In our scenario generation concept, we combine the best of both worlds: We explore scenario spaces by using a combinatorial approach for sampling discrete components and exploiting the parameter space by sampling continuous parameters. 
    
    \subsection{Semi-Concrete Scenarios}
    To combine exploration and exploitation in the context of variability modeling, we introduce \textit{semi-concrete} scenarios.
    A semi-concrete scenario combines logical and concrete scenarios by assigning concrete values for discrete parameters but leaves the value ranges of continuous ones. Consequently, generating semi-concrete scenarios that systematically cover all interactions of $t$ discrete scenario elements allows us to explore the scenario space systematically, but it leaves open the possibility to exploit each of the generated semi-concrete scenarios by sampling the continuous parameters. 
    To this end, we extend the existing scenario feature model proposed by Birkemeyer et al.~\cite{birkemeyer2022feature} to include semi-concrete scenarios. The novel scenario feature model structure can represent scenarios at different abstraction levels (i.e., logical, semi-concrete, and concrete). We refer to the resulting extended feature models as \textit{hybrid scenario feature models}.
    Figure~\ref{fig:logical_concrete_FM} exemplifies the hybrid scenario feature model.
    
    The \textit{hybrid scenario feature model} consists of (a) \textit{structure features}, (b) \textit{logical features}, and (c) \textit{concrete features} (cf. Figure~\ref{fig:logical_concrete_FM}).
    The \textit{structure features} are abstract and structure the scenario feature model in a tree structure, e.g., to allocate atomic scenario elements to the common six-scenario levels proposed in~\cite{scholtes20216}. \textit{Logical features} represent parameters and possible parameter ranges. In our example, (Vehicle-)\textit{Type} and \textit{Velocity} are logical features. The feature \textit{Type} includes a list of discrete parameters (i.e., $type=[A4, Beetle, X5, \ldots]$), whereas velocity is a continuous parameter (i.e., \mbox{$v=[0,210]~km/h$}). Selecting scenarios from a feature model that contains structure and logical features results in logical scenarios; thus, we define it as a \textit{logical feature model}. In Figure~\ref{fig:logical_concrete_FM}, we mark the \textit{logical feature model} with a blue bounding box. 
    
    Using a \textit{logical feature model}, we can add concrete features representing concrete parameter assignments and obtain concrete scenarios. For discrete parameters, such as the vehicle type, we add a concrete feature for each type; however, we need to discretize the range for continuous parameters. We do so by adding a feature for each equivalence class.
    Standard combinatorial approaches use expert-defined equivalence classes~\cite{birkemeyer2022feature, kluck2023empirical}. Adding concrete features to a logical feature model results in a \textit{concrete feature model}. The \textit{logical feature model} models the same logical scenarios as the remaining \textit{concrete feature model} (cf. Figure~\ref{fig:logical_concrete_FM}, orange bounding box). However, the \textit{logical feature model} implicitly models a broader space of concrete scenarios wrt. to continuous scenario parameters. 
    
    Finally, to select semi-concrete scenarios, we use the \textit{semi-concrete feature model} (cf. Figure~\ref{fig:logical_concrete_FM}, green bounding box). In this feature model, discrete parameter values are modeled as concrete features, but continuous parameters are represented with logical features.
    Sampling the continuous parameters can be done in various ways. For instance, one can sample concrete values for the continuous parameters within the entire range of possible values (see~Sect.~\ref{sec:parameter-sampling}) or within sub-parameter ranges (see~Sect.~\ref{sec:equivalence-classes}).
    
    \subsection{Parameter Range Sampling}
        \label{sec:parameter-sampling}
    
        During the modeling process of the semi-concrete feature models, we use logical features to represent representative upper and lower boundaries for continuous parameters. Those boundaries might be defined by expert knowledge or derived from real-world data. For example, the manufacturer specifies a vehicle's maximal velocity, whereas minimal and maximal rain intensity can be derived from real-world observations. These boundaries allow continuous parameters to be sampled over the entire range of values. In our evaluation, we opted to sample parameters randomly to avoid introducing any bias; however, continuous parameters could be sampled using optimization techniques or probability density functions.
        
        \subsection{Sub-Parameter Range Sampling}
        \label{sec:equivalence-classes}
            Similar to equivalence classes, we define \subrangepl that partitions the range of continuous parameter values into smaller ranges that are combined to cover each interaction of $t$ parameter ranges.
            In \subrange sampling, concrete values are sampled from those ranges every time a new concrete scenario is generated.
            Thus, in contrast to equivalence classes, \subrange sampling does not always select the predefined representative values defined by the domain expert. Nonetheless, it systematically covers (explores) parameter ranges.
            As for parameter range sampling, in our evaluation, we opted to sample parameters within each \subrange randomly. Continuous parameters could be sampled using optimization techniques or probability density functions with or against a distribution.
            
            Adopting \subrange sampling requires (1) generating semi-concrete scenarios and (2) sampling concrete values for each of them. Parameter range sampling, instead, does not require covering each interaction of $t$ parameter ranges; consequently, it is less expensive, as the parameters are sampled fewer times. 
            However, because the value ranges considered by parameter range sampling are larger than the ones considered by \subrange sampling, the generated test cases might be less effective. 
            In summary, parameter range sampling only exploits the continuous parameters defined from the explored logical scenarios, whereas \subrange sampling adds an intermediate level of exploration, i.e., exploring the interaction of $t$ sub-parameter ranges, before the final parameters exploitation. By selecting one or the other option, developers can trade off the cost and effectiveness of the generated test suites.
        
\section{Evaluation}
    To assess the benefits of combining combinatorial testing and parameter sampling and to understand how parameter sampling affects the effectiveness of the generated test scenarios, we investigate the following main research questions:

    \begin{enumerate}[label=\textbf{RQ\arabic*}]
        \item \emph{Does combining combinatorial scenario generation and parameter sampling generate more effective test suites than standard combinatorial scenario generation?}
        Parameter sampling enables exploiting continuous parameters. Hence, in the sense of fault detection ability, we want to understand if this leads to generating more effective test suites than using expert-defined~values.
        
        \item \emph{How does a specific sampling technique impact test suite effectiveness?}
        Sampling parameters using \subrange sampling is more expensive than parameter range sampling in the overall possible range. This raises the question of whether \subrange sampling generates test suites that identify more faults than the one generated by parameter range sampling.
        
        \item  \emph{How does ``sampling with feedback'' impact the effectiveness of a test suite?}
        Optimization techniques use a feedback loop to guide the parameter sampling process. We are interested 
        if parameter sampling \textit{with feedback} outperforms purely random parameter sampling when combined with combinatorial scenario generation.
    
    \end{enumerate}    

    \begin{figure*}[t]
        \centering
         \subfloat[feature-wise interaction coverage YASA (t=1)]
        {\includegraphics[width=0.49\linewidth]{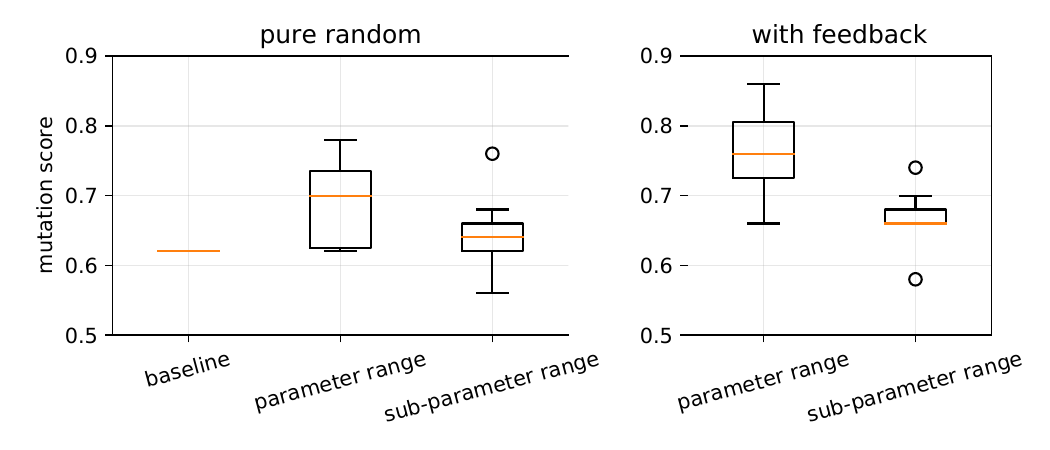}}
        \hspace{-0.8cm}
        \qquad
        \subfloat[pair-wise feature interaction coverage YASA (t=2)]
        {\includegraphics[width=0.49\linewidth]{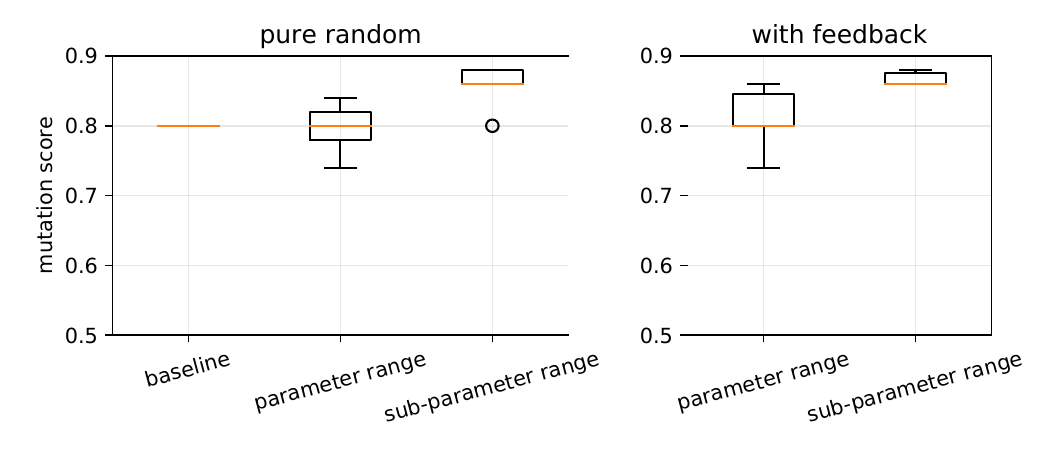}}
        \vspace{-0.3\baselineskip}
        \caption{Test effectiveness of scenario suites generated using combinatorial scenario generation and parameter sampling for YASA \textit{t=1} (a) and YASA \textit{t=2} (b). We separate parameter sampling strategies that select parameters purely randomly or with an iterative feedback loop. For readability, we removed outliers that lead to mutation scores smaller than 0.5
        \vspace{-0.3cm}
        }
        \label{fig:results_mutation_score}
    \end{figure*}
            
    \subsection{Experimental Setup}
    To evaluate the effectiveness of combining combinatorial scenario generation and parameter sampling,
    we implemented the proposed approach in a prototype\footnote{\url{https://doi.org/10.5281/zenodo.7940902}\label{ftn:zenodo}} and used it to generate concrete scenarios. 
    We executed the concrete scenarios in the virtual environment provided by IPG Carmaker version 11.0.1.\footnote{\label{note:ipg}\url{https://ipg-automotive.com/}}
    As a test subject, we implemented an adaption of the \acrfull{ACC} provided by IPG and Mathworks\footnote{\label{note:mathworks}\url{https://mathworks.com/}} in Simulink. The ACC adapts the ego vehicle's speed to the leading vehicle's speed.
    Following an iterative process that involved two researchers, we implemented a hybrid feature model representing relevant scenarios to test \gls{ACC} functionality.
    The resulting hybrid feature model contains $193$ features in total, with $31$ parameter features and $93$ expert-defined sub-parameter ranges, i.e., three \subrangepl per continuous parameter.
    We share the feature model online.\footref{ftn:zenodo}
    Using the hybrid feature model and our prototype, we generated test suites by sampling semi-concrete scenarios with parameter ranges and sub-parameter ranges. In both cases, we used the state-of-the-art coverage-based sampling algorithm YASA~\cite{krieter2020yasa} to cover the interaction of $t$ parameters; we used YASA \textit{t=1} (feature-wise) and YASA \textit{t=2} (pair-wise).
    Specifically, \textit{parameter ranges} produced 5 (YASA \textit{t=1}) and 30 (YASA \textit{t=2}) semi-concrete scenarios, 
    whereas \textit{\subrangepl} produced $7$ (YASA \textit{t=1}) and 80 (YASA \textit{t=2}) semi-concrete scenarios.
    To transfer semi-concrete scenarios into concrete scenarios, we used randomized parameter sampling, which we repeated $10$ times to increase confidence in our conclusions.
    As a \cready{\textit{baseline}} for comparison, we considered the \subrangepl as equivalence classes and selected scenarios with state-of-the-art combinatorial scenario generation. We generated test suites by using YASA (\textit{t=1} and \textit{t=2}) and assigned the expert-defined values to all the equivalence classes (i.e., sub-parameter ranges) instead of sampling them.
    Notably, since the baseline uses the same \subrangepl used by \subrange sampling, it produced the same amount of concrete scenarios, i.e., $7$ (YASA \textit{t=1}) and $80$ (YASA \textit{t=2}) concrete scenarios.
    Specifically,
    for each comparison, we measure \emph{statistical significance} (p-value) using the Mann–Whitney U-test. Broadly speaking, p-values less than $0.05$ indicate statistically significant results. 
    In addition to assessing statistical significance, we measure the \emph{effect size} using the Vargha and Delaney A$_{12}$, which captures the magnitude of the differences between two distributions (i.e., the effectiveness of two scenario suites). 
    For example, a value of A$_{12}$ equal to $=0.5$ means that both suites are equally effective; thus, the difference in their effectiveness is \emph{negligible}. However, as the A$_{12}$ value moves away from the $0.5$ threshold, the difference in test suite effectiveness becomes \emph{small}, \emph{medium}, or \emph{large}.

    \textit{Feedback Loop:}
    Regarding RQ~3, we guide parameter sampling with a feedback loop trying to increase the test suites' effectiveness by limiting the occurrence of potentially irrelevant scenarios.
    \cready{The feedback loop is a first step and considered as a proof of concept in combining coverage-based sampling (exploration) and optimization techniques (exploitation).}
    We introduce an independent \acrfull{SRC} to determine whether a scenario is relevant. Regarding the \gls{ACC} as \gls{SUT}, we define a concrete scenario as relevant if at least one object occurs within the sensor's field of view within the first $10$ seconds. We re-sample parameters up to $50$ times if a scenario is irrelevant.
    
    \textit{Test Suite Effectiveness:}
    Inspired by Birkemeyer et al.~\cite{birkemeyer2022feature}, we assessed the test effectiveness of all generated test suites using fault-based testing~\cite{57623}. Thus, to assess the ability of tests to expose problems in the \gls{SUT}, we purposely injected faults in it and checked whether the tests identified them. We used the \textsc{SIMULTATE} Framework~\cite{pill2016simultate} to generate $50$ \cready{random} faulty versions of the \gls{ACC} and counted the ones that are detected by the test suite.
    Intuitively, the more faults are found by a test suite, the more effective that test suite is. Therefore, as an effectiveness metric, we use the \textit{mutation score}, i.e., the ratio of detected faulty versions to all.

    \subsection{Results and Discussion}
        
    \paragraph*{RQ1. Benefits of Combining Combinatorial Scenario Generation and Parameter Sampling} 
        Figure~\ref{fig:results_mutation_score} reports the number of faults found, i.e., \emph{mutation score}, by each test suite we generated using YASA \textit{t=1} (Figure~\ref{fig:results_mutation_score}-a) and YASA \textit{t=2} (Figure~\ref{fig:results_mutation_score}-b).
        From the figure, we can observe that for YASA \textit{t=1}, both sampling strategies generate more effective test suites than the baseline.
        Comparing the baseline and Parameter Range Sampling resulted in a statistically significant result (p-value$=0.042$) and a \emph{large} effect size while comparing the baseline and \Subrange Sampling resulted in a less statistically significant result (p-value$<0.069$) and \emph{medium} effect size. 
        For YASA \textit{t=2}, we can observe that the baseline and Parameter Range Sampling achieved comparable results (p-value=$1.0$ and negligible effect size); however, \Subrange Sampling generates statistically more effective test suites (p-value$<0.005$ and \emph{large} effect size).
        Finally, comparing the results achieved by covering single features (YASA \textit{t=1}) and pair-wise interactions between them (YASA \textit{t=2}), we can observe that -as expected- generating more scenarios lead to more effective test suites.

        In light of those observations, we conclude that combining combinatorial scenario generation and parameter sampling is beneficial for scenario-based testing as it 
        systematically covers all interactions of $t$ features while outperforming basic combinatorial strategies discretizing continuous parameters.

    \paragraph*{RQ2. Impact of Sampling on Test Effectiveness}
        To address RQ2, we refer again to 
        Figure~\ref{fig:results_mutation_score} and compare the results obtained using Parameter Range Sampling and \Subrange Sampling.
        We observe that none of the two methods always produces the most effective test suite. Specifically, for YASA \textit{t=1}, Parameter Range Sampling produced a slightly better test suite (p-value$=0.3$ and \emph{small} effect size). On the contrary, for YASA \textit{t=2}, \Subrange Sampling produced more effective test suites (p-value=$<0.005$ and \emph{large} effect size).
        Although Parameter Range Sampling generated smaller test suites, it achieved good results (i.e., never worse than the baseline).

        In light of those observations, we conclude that Parameter Range Sampling is a viable choice when scenario generation focuses on covering all scenario features or when scenario-based testing is budget-constrained, while \Subrange Sampling might be a better choice when developers aim to fulfill stronger coverage criteria and can afford to generate larger test suites.
        
    \paragraph*{RQ3: Impact of Sampling with Feedback on Test Effectiveness}
        To answer RQ3, we compare the test effectiveness for scenario suites generated \textit{with} and \textit{without} feedback loop during the parameter sampling process. The results are shown in Figure~\ref{fig:results_mutation_score} ("pure random" vs. "with feedback").
        We assume that the test effectiveness increases for parameter sampling \textit{with} feedback. Regarding YASA \textit{t=1}, there is a clear increase of effectiveness for both sampling strategies: Parameter Range Sampling (p-value$=0.03$ and \emph{large} effect size) and \Subrange Sampling (p-value$=0.4$ and \emph{small} effect size). Regarding YASA \textit{t=2}, however, the increase is marginal; (p-value=$0.5$ and \emph{small} effect size) for Parameter Range Sampling and (p-value=$0.9$ and \emph{negligible} effect size) for \Subrange Sampling.
        It is worth mentioning that both parameter sampling strategies
        \textit{with} feedback perform similarly to each other for YASA \textit{t=1} and \textit{t=2} as both strategies \textit{without} feedback: Parameter Range Sampling leads to more effective scenario suites for YASA \textit{t=1}, while \Subrange sampling is more effective for YASA \textit{t=2} (cf. RQ~2).

        In light of those observations, we conclude that adding feedback to parameter sampling slightly increases the test effectiveness of generated scenario suites. However, during our experiments, parameter sampling with feedback requires in median of 1.2\% (overall parameter range) and 14\% (sub-parameter range) additional simulation effort. For the sake of fairness in this comparison, we use simple resource allocation (50 tries for each (sub-)parameter range)~\cite{campos2014continuous}. But, relevant parameters might be rare in some (sub-)parameter ranges, which impacts the additional effort.
        However, depending on the use-case, test engineers need to carefully weigh costs and benefits while generating \gls{SOTIF}-compliant scenario suites.


\subsection{Threats to Validity}
    
    \paragraph{Internal validity} 
    A threat to the internal validity of our study is relying on expert knowledge to define the feature model, equivalence classes, etc. To mitigate this threat, we rely on the knowledge of two independent experts. However, our results show that using the values defined by equivalence classes does not lead to equivalent behavior in all scenarios. We argue that the industry will face similar challenges; thus, our results are \cready{meaningful}.
    Another threat comes with the definition of \textit{relevant scenarios} for the feedback loop. We argue that the definition is arbitrary and a systematic study in the future could address this threat.
    Finally, a threat results from the (external) tools we used for evaluation. To mitigate this threat, we rely on established tools (Carmaker and Simulink) in the automotive industry and adapted an existing implementation of \gls{ACC}.
             
    \paragraph{External validity} 
    Considering only \gls{ACC} functionality in the evaluation cannot let us generalize the conclusions to other \gls{ADAS} or \gls{ADS}.
    \cready{However, we argue that \pgls{ACC} is established in modern vehicles; hence, it must be tested.}
    
\section{CONCLUSION}
    This paper establishes a novel concept for semi-concrete scenario generation and parameter sampling to generate effective test suites for scenario-based testing of \gls{ADAS}/\gls{ADS}. The proposed concept allows developers to balance the trade-off between exploring and exploiting driving scenario spaces as intended by \gls{SOTIF} (ISO 21448). Based on our results, we conclude that semi-concrete scenario generation and parameter sampling increases the effectiveness of \gls{SOTIF}-compliant scenario generation to verify and validate \gls{ADAS}/\gls{ADS}. Moreover, depending on the parameter sampling strategy (full-/sub-parameter ranges, with/without feedback), our results indicate the potential for different testing strategies.
    In the future, we aim to improve the test effectiveness of the generated scenario suites and address the major external and internal threats to validity. We will add real-world data to define continuous parameter ranges and focus on strategies to define sub-parameter ranges automatically. Moreover, we will apply optimization techniques such as search-based or stochastic optimization to improve parameter sampling with feedback to balance resource allocations.
    
    \section*{Acknowledgments}
    We thank David Schultz for constructive discussions and for supporting the implementations. This work has been partially funded by the Ph.D. program "Responsible AI in the Digital Society" funded by the Ministry for Science and Culture of Lower Saxony, Germany, and the 
    EU Project Flexcrash (Grant Agreement n. 101069674).
    This work has been partially supported by projects within the Innovations Campus Future Mobility (ICM) funded by the Ministry of Science, Research and Arts of the Federal State of Baden-Württemberg.


\addtolength{\textheight}{-10cm}   


\balance
\bibliographystyle{IEEEtran}
\bibliography{bib/bib.bib}

\end{document}